\documentstyle [floats,prl,aps]{revtex}
\input epsf
\begin{document}

\draft
\title{A new approach to neutrino and WIMP detection \\
using
telecom-grade electrooptic and fiber-optic components}

\author{
J.I. Collar$^{*}$
}
\address{
Groupe de Physique des Solides (UMR CNRS 75-88), Universit\'es Paris 
7 \& 6, 
75251 Paris Cedex 05, France\\
$^{*}${\it present address:} Enrico Fermi Institute,
5640 S. Ellis Ave.,
Chicago, IL 60637, USA\\
}
\wideabs{
\maketitle
\begin{abstract}
\widetext
The dense energy deposition from a low-energy 
nuclear recoil produces, via the thermoacoustic effect, a 
brief yet intense pressure pulse that can be used for WIMP or neutrino 
detection in some fiber-optic devices sensitive to acoustooptic 
disturbances. Several possible detection schemes are described: all of 
them are inspired by modern fiber-optic sensor 
technologies and
share common characteristics of 
low-cost and expected insensitivity to minimum-ionizing backgrounds. 
\end{abstract}

\pacs{ \\
}}
\narrowtext

The detection of low-energy nuclear recoils
originating in neutrino or
Weakly Interacting Massive Particle (WIMP) elastic scattering off
nuclei has been the subject of a great experimental effort in the last
two decades. The difficulties involved in the development of detectors
adequate
for this task are multiple: {\it i}) the signal rate expected is generally
very low (few counts per kg target per day or much less), imposing the
need for large (multi-kg) detector masses, {\it ii}) the signal accumulates
in the keV or sub-keV region, requiring a low energy threshold, {\it
iii}) the recoil signal must be extruded from a sea of dominant backgrounds,
most of them originating in minimum-ionizing particles. The interest
in this particular area of detector development is amply justified: WIMPs
constitute one of the best candidates for Dark Matter
at the galactic level \cite{WIMPreview}. Numerous ongoing and planned searches
aim at the detection of WIMP-induced recoils\cite{dan}, employing a variety
of techniques. Some of the most sensitive present
WIMP detectors rely on costly and cumbersome cryogenic techniques.
Unfortunately,
some theoretically well-motivated WIMP
candidates such as those arising from supersymmetric extensions of
the Standard Model can have elastic scattering cross sections so small that
only ton
or multi-ton background-insensitive devices have a realistic chance at
their direct detection. Low-cost and simplicity become in this way a
determinant factor in the design of next-generation WIMP detectors.

A second important objective for these devices is the detection of
{\it coherent}
neutrino scattering:
an uncontroversial process in the Standard Model, the scattering off
nuclei of low-energy neutrinos ($< $ few tens of MeV) via the neutral current
\cite{freedman} remains undetected. The long neutrino wavelength probes
the entire nucleus, giving rise to a large coherent enhancement in the cross
section, roughly proportional to neutron number squared
\cite{drukier}. A quantum-mechanical condition
for the appearance of coherent effects is the
indistinguishability of initial and final states and hence the absence of a
charged-current equivalent. In principle, it would be possible to speak of
{\it portable} neutrino detectors since the expected rates can be as high
as several
hundred recoils/kg/day (\frenchspacing{Fig. 1}), by no means a ``rare-event''
situation. A detector of this type would open the door to extraordinary
applications in ``neutrino technology''\cite{leo,leo2} (planetary tomography
and prospecting, telecommunications,
extra-galactic neutrino detection and strategic applications).
However, the recoil energy transferred to the target is of a few keV at
most for
the lightest nuclei (\frenchspacing{Fig. 1}): several ambitious
proposals to use
new-generation cryogenic
detectors \cite{drukier,blas,starostin} have been put forward, but {\it no
existing
device} meets the target mass and energy threshold requirements involved in
this
measurement. The interest in observing this process is not purely
academic nor the stuff of science-fiction. For instance, a
neutral-current detector responds the same way to all known neutrino
types,
meaning that the observation of neutrino oscillations in such a device would
be direct evidence for a fourth sterile neutrino. These must be invoked if all
recently observed neutrino anomalies are accepted at face value
\cite{john} and may play an important role as Dark Matter \cite{dolgov}.
Separately, the cross section for this process is critically dependent on
neutrino magnetic moment: concordance with the Standard Model prediction
would {\it per se} largely improve the present experimental sensitivity
to $\mu_{\nu}$\cite{dodd}.
Finally, this mechanism plays a most important role in neutrino
dynamics in supernovae and neutron stars \cite{freedman}, adding to the
attraction of a laboratory measurement.
\begin{figure}[tbp]
\epsfxsize = \hsize \epsfbox{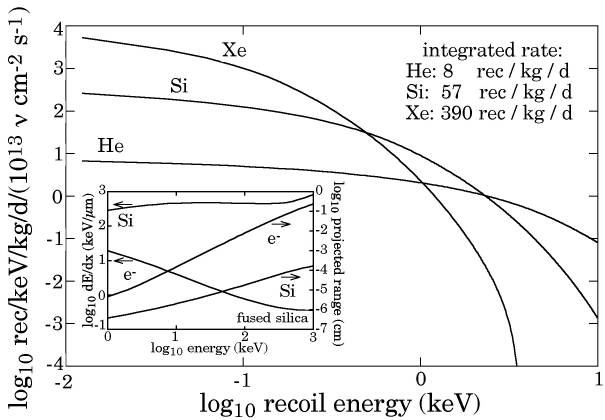}
\caption{Nuclear recoil energy spectrum from
neutral-current scattering of reactor
neutrinos on different targets.
{\it Inset}: stopping power and range of low-energy Si recoils and
electrons in fused silica (SiO$_{2}$).}
\end{figure}

The research project described next has as its goal the development of
cost-effective cryogenic-free detector technologies responsive to these
low-energy
recoils while being insensitive to competing backgrounds.
The innovative approach proposed
profits from the enormous ongoing progress in optoelectronic components
and devices,
making use of inexpensive telecom-grade optical fibers,
light sensors and amplifiers.
\newpage

At the microscopic level (nm scale) a nuclear recoil is an extremely
violent event. The scenario, common to other forms of high-LET particle
interaction, starts with the formation of a highly-ionized plasma
in the recoil aftertrack: for sufficiently dense energy depositions
the nuclear and electronic temperatures can
reach local values in the thousands of degrees Kelvin. This
transient ($\sim\!10^{-12}$ s) heat pulse or ``hot-spike''
\cite{seitz} is soon quenched: the rapid adiabatic expansion of the
material provokes a considerable shock wave (the ``thermoacoustic''
effect \cite{acousto}), reaching local peak pressures in excess of $10^{5}$
atm \cite{sun} that take away a large
fraction of the deposited energy in the form of phonons. This
acoustic shock-wave has been used before as a monitoring method
for intense beams of minimum-ionizing radiation \cite{learned,aska}
and can play a role in radiation damage to tissue \cite{sun}.
The frequency of the
dominant sound emission is $\sim c_{s}/l$ ($c_{s}$ is the sound
speed, $l$ is the transverse track size), which for keV energy
recoils,
having a comparable $l$ and range (\frenchspacing{Fig. 1,
inset}), is in the tens or hundreds of GHz. This is
unfortunate, since such frequencies are rapidly damped in most room-temperature
materials
over distances of the order of few microns \cite{sun,aska}. This
limitation is bypassed in
a proposal to use Si crystal bolometers, where ballistic phonons can
propagate and be detected
over macroscopic (cm) distances at cryogenic temperatures
\cite{blas}.
While derivatives of this approach have found successful applications in
WIMP searches
\cite{cdms}, the feasibility of scaling it up to the large target
masses required for the ultimate WIMP detector remains an open
question. These bolometers currently display effective energy thresholds
still far too
high for
coherent neutrino detection, with no immediate improvement in sight.

The central question is then: can a brief ($\sim$ns) yet intense pressure
pulse
propagating over a short ($\sim\!\mu$m) distance in an room-temperature
device be used for efficient nuclear recoil detection? This proposal intends
to show that for some unsophisticated contraptions the answer may be ``yes''.

A brief exercise in mental gymnastics provides a first approach:
take the O(kg) target mass, elongate it until its
geometrical cross section becomes comparable to the size of the
recoil-induced disturbance and send a probe through it, one able to
carry the fleeting information of this short-lived event to a
monitoring instrument down the line. To envision a fiber-optic device,
where the transmitted light is the probe and the fiber itself the
target is then only natural. In an era when fiber-optic
communications are becoming inexpensive, commonplace and transmission
speeds are rapidly approaching the THz barrier, the idea seems
timely. What is more, a plethora of fiber-optic sensing devices
\cite{udd,bucaro} have amply demonstrated the exquisite sensitivity that
they provide in a variety of applications.

\begin{figure}[tbp]
\epsfxsize = \hsize \epsfbox{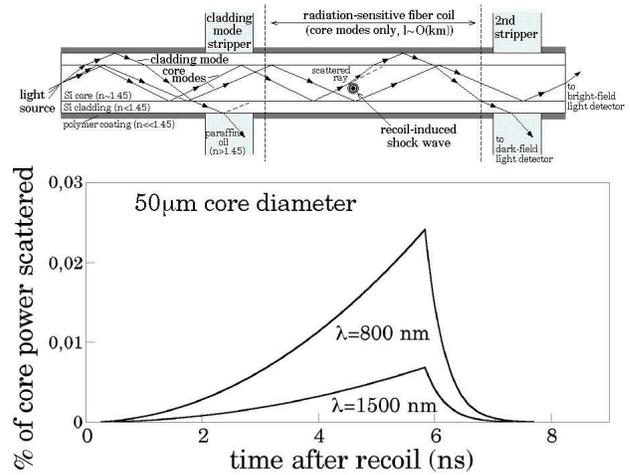}
\caption{{\it Top}: conceptual design of a low-energy recoil detector
using HCS telecom fiber. {\it Bottom}: estimated magnitude of a
keV recoil-induced disturbance using
Mie scattering theory.}
\end{figure}
\frenchspacing{Fig. 2 (top)} illustrates a first example. The strategy
depicted there is freely inspired by the ``microbend'' acoustic fiber
sensor \cite{bucaro,lagakos}. A common multimode hard-clad
silica (``HCS'') fiber, of wide use in industrial control applications
presents the interesting property of being a true dual lightguide:
injected photons can propagate either through core modes or external
higher-order cladding modes (\frenchspacing{Fig. 2, top}). These cladding
modes can be completely removed by immersing a short
segment of denuded fiber in a
high-index of refraction ($n$) liquid (a so-called ``mode stripper''),
``cleaning'' one of the two possible light
paths. While some mixing of core and cladding modes (``mode beating'')
can occur down the
fiber due to
careless winding of the fiber coil, scattering on microbubbles, etc., this
separation can be maintained over long distances (a first prototype under
construction by the author at Groupe de Physique des Solides
displays a leakage of the core-transmitted power
to the cladding of $<3\%$ over a tight 100 m coil). In a microbend acoustic
sensor, a special mechanical transducer excited by acoustic pressure can
affect
the core modes in such a way that they feed light again to the otherwise
blank cladding channel (present microbend sensors are sensitive to 
transducer displacements of less than an Angstrom). 
This cladding light can then be extracted at
the end of the sensing fiber by a second mode stripper and detected, providing
a measure of the acoustic signal \cite{lagakos}. Here, the role
of the transducer can be played instead by the short-lived changes in
density, polarizability and index of refraction
expected to accompany a recoil-induced shock-wave (i.e., the
well-known ``acoustooptical''
effect \cite{yariv,flax}), which are able to deflect (even if minimally)
some of
the core power into higher modes (\frenchspacing{Fig. 2, top}). The
duration and intensity of the cladding pulse so produced can be
calculated using Mie scattering theory (\frenchspacing{Fig. 2, bottom)}
as is generally done to estimate the
effect on light attenuation of micron-sized density fluctuations in
fibers
\cite{mie}. The
calculation accounts for the
magnitude and propagation of the pressure wave
following \cite{sun,learned,aska}, includes its
large attenuation
at high frequencies
\cite{pelous}, uses the index of refraction of shock-compressed
fused silica as in \cite{set} and pays attention to the fact that
shock-wave and sound speeds are not the same in some 
regimes \cite{http}. To give the reader a reference point, the $t\!=\!0$
change in $n$ at the recoil site can be as large as $\sim 7\%$.
While small, the  signal expected from this design
can in principle be easily detected with fast photodiodes like those
now commonly used in optical telecom networks and amplified with
low-noise chips like those developed for GHz-band cellular
telephony. The first prototype device presently under development is
expected to have a sensitivity at least two orders of magnitude
better than what is predicted by the Mie estimate (in order to
accomplish this,
several features such as a low-noise, high-intensity LED light source,
selective mode injection \cite{jeu}, careful choice of PIN plus amplifier,
and a $4\pi$
cladding light collector are being implemented).

Several remarks are in order: the bandwidth and attenuation properties
of the fiber must allow for this brief signal to propagate over the long
fiber distances needed to ensure a large-mass recoil
detector. The excellent quality of modern telecom fiber
guarantees this. For the same reason (substantial target mass),
only large core-diameter ($\geq 100~\mu$m) multimode
fibers seem to be of interest for the goal in mind. Specialty
fiber (polarization-preserving, single-mode, etc.) defeats the purpose
by being not only much more
expensive but having an insufficient mass-to-length ratio. Most
importantly, the devices
envisioned are expected to be largely insensitive to
isolated minimum-ionizing radiation, the reason being the large dependence
of the magnitude of the thermoacoustic pressure spike on
particle stopping power \cite{sun,learned,aska}:
the sparse energy deposition typical of low-LET particles
(\frenchspacing{Fig. 1, inset})
is expected
to produce no measurable effect. The fact that the
fused silica used in commercial fibers is of low intrinsic
U and Th-chain radioactivity is of special mention: even small
concentrations of metal impurities
are known to
jeopardize light
transmission and hence a special industrial 
process of silica preform purification is
followed.
Therefore, a very low rate of events arising from
alpha-recoils or fission fragments in the fiber material is
anticipated.
\begin{figure}[tbp]
\epsfxsize = \hsize \epsfbox{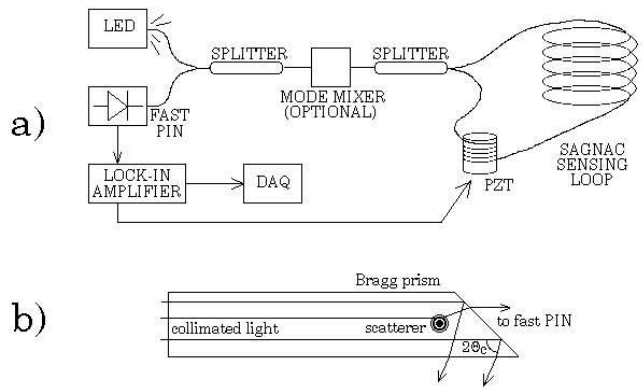}
\caption{{\it a)} A minimal-configuration Sagnac fiber optic interferometer
using multimode fiber. {\it b)} Can the combination of a Bragg prism and a
fast PIN diode be used as a sensitive (dark-field)
recoil detector? }
\end{figure}

Many of the principles used in other
fiber-optic sensors can be adapted for the detection of WIMP or
neutrino recoils. For instance, another possibility is the use of
a Sagnac interferometer \cite{udd,udd2}, a design widely adopted in
commercial fiber-optic
gyroscopes for its simplicity and robustness vis-a-vis external
disruptions (temperature gradients, sound, etc.). A more
in-detail look at the interaction process described above would show that the
shock-wave must produce a strong iridescence effect
on the light it intersects, due to the comparable photon wavelength and
shock-train spatial width. The effect of this phase shift on a Sagnac sensing
loop (which can be up to few km long) should be readily measurable,
provided that the ``gyroscope'' is
instrumented to look for nanosecond phase changes, which is evidently
never the case in normal use. An exception to this statement is found in new
ultrafast optical switches such as the Sagnac TOAD \cite{toad}, which
presents a strong parallel to the recoil detection scheme outlined here
(the curious reader is encouraged to consult \cite{toad}).
Interestingly enough, minimum-configuration,
low-cost fiber optic gyros can be assembled out of the thick
multimode fiber required for recoil detection \cite{fredi}. Other
schemes using polarization measurements and Bragg prisms are under
study (\frenchspacing{Fig. 3}). The simultaneous use of multiple
photon wavelengths
for redundancy and noise
rejection can be envisioned (wavelength-selective
emitters, sensors and splitters are now common use in telecom).

After a working prototype is achieved, calibration in a variety of
neutron fields would be the next natural step. Needless to say, what
is good for neutrino recoils is good for the detection of neutrons.
Present fiber optic dosimeters are rudimentary \cite{dosim}
in the sense that they offer no
real-time detection of single particles. Passive scintillating fibers
are an exception, however of no use here for reasons of cost,
absorption length
(which limits target mass), lack of built-in background discrimination
and high energy threshold. Therefore, applications in neutron dosimetry should
not be discounted. 
A challenge will be the extraction of energy
information from the devices (even though most of the applications
described in the first part of this paper can be pursued with
threshold detectors): time-frequency and amplitude analysis of the signals
can cast light on this.

In conclusion, the tools necessary for novel low-cost,
room-temperature, background-insensitive neutrino and WIMP detectors
seem to be up for grabs, thanks to the extraordinary recent developments in
the at first sight unrelated fields of telecom, optoelectronics and fiber
optic sensors.

{\it Acknowledgements:} I am indebted to Jim Wolfe for suggesting 
the possibility of using a Bragg prism.

\newpage

\end{document}